\documentclass[10pt]{article}
\usepackage{fleqn,espcrc1}
\usepackage{graphicx}
\usepackage[figuresright]{rotating}

\newcommand{\AmS}{{\protect\the\textfont2
  A\kern-.1667em\lower.5ex\hbox{M}\kern-.125emS}}
\title{Neutrino Processes in Supernovae 
and the Physics of Protoneutron Star Winds}
\author{Todd A. Thompson 
	\address{Department of Physics, The University of Arizona, Tucson 85721}
        \thanks{Electronic Address: thomp@physics.arizona.edu} and
        Adam Burrows \address{Steward Observatory and Department of Astronomy,
The University of Arizona, Tucson 85721}
\thanks{Electronic Address: burrows@jupiter.as.arizona.edu}}
       
\begin{document}
\maketitle

\begin{abstract}
In preparation for a set of hydrodynamical simulations of 
core-collapse supernovae and protoneutron star winds, we 
investigate the rates of production and thermalization of 
$\nu_\mu$ and $\nu_\tau$ neutrinos in dense nuclear matter.  
Included are contributions from electron scattering, 
electron-positron annihilation, nucleon-nucleon bremsstrahlung, 
and nucleon scattering. We find that nucleon scattering dominates 
electron scattering as a thermalization process at neutrino energies 
greater than $\sim$15 MeV.  In addition, nucleon-nucleon 
bremsstrahlung dominates electron-positron annihilation 
as a production mechanism at low neutrino energies, near and below 
the $\nu_\mu$ and $\nu_\tau$ neutrinosphere.

Furthermore, we have begun a study of steady-state general relativistic 
protoneutron star winds employing simple neutrino heating and
cooling terms.  From this analysis we obtain acceleration
profiles as well as asymptotic lepton fractions and 
baryon entropies essential in assessing the wind
as a potential site for $r$-process nucleosynthesis.

\end{abstract}

\section{Introduction}
\label{sec:intro}
The cores of protoneutron stars and core-collapse supernovae are unique 
environments in nature. They are characterized by mass densities of order 
$\sim 10^{11}-10^{15}$ g cm$^{-3}$  and temperatures that range from 
$\sim 1$ to $50$ MeV.  At these temperatures and densities neutrinos of 
all species are produced in proliferation via electron-positron annihilation 
($e^+e^-\leftrightarrow \nu\bar{\nu}$), nucleon-nucleon bremsstrahlung, 
and plasmon decay ($\gamma_{pl}\leftrightarrow\nu\bar{\nu}$).  While these 
processes contribute for the electron types ($\nu_e$s and $\bar{\nu}_e$s), 
for them the charged-current absorption and emission processes 
$\nu_e n\leftrightarrow p e^-$ and $\bar{\nu}_ep\leftrightarrow ne^+$ dominate. 
Neutrino-electron and neutrino-nucleon scattering also contribute to the total 
opacity.  All of these interactions combine to couple the neutrinos to dense 
nuclear matter, affecting energy transport from the core where the neutrinos 
are diffusive to the more tenuous outer layers where the neutrinos begin 
to free-stream.  Indeed, the neutrino heating in the semi-transparent region 
behind the shock is now thought to be an essential ingredient in igniting 
the supernova explosion itself \cite{colgate,bethe,bhf_1995}.  Furthermore, 
a neutrino-driven protoneutron star wind is thought to be a general feature 
of the core-collapse phenomenon and has been proposed as a site for $r$-process 
nucleosynthesis.

Essential in understanding the mechanism of Type-II supernovae is an accurate 
prediction of the production spectrum for each neutrino species.  To this end, 
in \S 2 we present results of a thermalization and equilibration study of 
$\mu$ and $\tau$ neutrinos in dense matter. 

Separately, in \S 3 we present preliminary results from calculations of 
steady-state protoneutron star winds in general relativity.  
We include velocity profiles, asymptotic entropies, and expansion timescales 
central in assessing this site as a candidate for $r$-process nucleosynthesis.

%-----------------------------------------------------------------------------------%

\section{Thermalization and Production}

For $\nu_\mu$ and $\nu_\tau$ neutrino types (collectively `$\nu_\mu$s'), which 
carry away  50$-$60\% of the $\sim 2-3\times 10^{53}$ ergs liberated during collapse 
and explosion, the prevailing opacity and production sources are $\nu_\mu$-electron 
scattering, $\nu_\mu$-nucleon scattering, $e^+e^-$ annihilation, and nucleon-nucleon 
bremsstrahlung.  The charged-current reactions dominate the electron-type transport 
so completely that we do not consider them here.  Our focus is on (1) the role of 
$\nu_\mu$-nucleon scattering relative to $\nu_\mu$-electron scattering in thermalizing 
$\nu_\mu$ neutrinos and (2) the importance of bremsstrahlung 
as compared with $e^+e^-$ annihilation as a producer of $\nu_\mu\bar{\nu}_\mu$ pairs.

Supernova theorists had long held \cite{lamb_pethick} that $\nu_\mu$-nucleon scattering did
not aid in thermalizing any neutrino species.  While the process was included as a source of 
opacity \cite{bhf_1995,bruenn_1985} it served only to redistribute the neutrinos in space, 
not in energy. In contrast, $\nu_\mu$-electron scattering was thought to dominate as a 
thermalizer of $\nu_\mu$s.  In addition, the only $\nu_\mu\bar{\nu}_\mu$ pair production 
mechanisms employed in full supernova calculations were 
$e^+e^-\leftrightarrow \nu_\mu\bar{\nu}_\mu$ and plasmon decay 
($\gamma_{\rm pl}\leftrightarrow \nu_\mu\bar{\nu}_\mu$) \cite{bruenn_1985}. 
Nucleon-nucleon bremsstrahlung, while recognized as a late-time cooling mechanism 
for more mature neutron stars \cite{fsb_1975,friman}, was neglected in supernova theory.  
Recent developments, however, 
call both these practices into question and motivate a re-evaluation of these processes 
in the supernova context. In the last few years, analytic formulae for $\nu_\mu$-nucleon 
have been derived  that include the full kinematics and final-state Pauli blocking, at 
arbitrary nucleon degeneracy, at the temperatures and densities encountered in the core 
of a core-collapse supernova \cite{reddy_1998,burrows_sawyer,bs_1999,reddy_1999}.  These 
efforts reveal that the average energy transfer in $\nu_\mu$-nucleon scattering may surpass 
previous estimates by an order of magnitude and, hence, that this process may compete with 
$\nu_\mu$-electron scattering as an equilibration mechanism
\cite{keil_1995,janka_1996,raffelt_seckel,sigl_1997}.  
Similarly, estimates of the nucleon-nucleon bremsstrahlung rate have been obtained  
\cite{fsb_1975,friman,burrows_1999,hannestad} which indicate that this process can compete with  
$e^+e^-$ annihilation in the dense core.

In order to compare these scattering and production processes directly,
we solve the Boltzmann equation for the time evolution of the neutrino phase-space
distribution function (${\cal F}_\nu$) in an idealized system with no spatial or 
angular gradients.  We consider an isotropic homogeneous thermal bath of scatterers
and absorbers held at constant temperature, density, and electron fraction.  
For the scattering processes, we begin
the equilibration calculation at $t=0$ with a $\nu_\mu$ distribution function with 
a characteristic temperature of twice that of the surrounding  matter.  We then evolve
this distribution function using the full collision integral of the Boltzmann equation,
with the structure function formalism of Burrows and Sawyer (1998) \cite{burrows_sawyer} 
and Reddy {\it et al.} (1998) \cite{reddy_1998}.  
When equilibrium is reached the final distribution is Fermi-Dirac, at the temperature
of the surrounding matter, with chemical potential set by the initial neutrino number density ($n_\nu$),
conserved to better than 0.0001\% throughout the calculation.
For the production and absorption processes we begin with zero phase-space occupancy
for both $\nu_\mu$ and $\bar{\nu}_\mu$ at all energies and let bremsstrahlung and $e^+e^-$
annihilation each build an equilibrium distribution of neutrinos and anti-neutrinos.
$e^+e^-$ annihilation is calculated in a Legendre polynomial expansion \cite{bruenn_1985}.  The production rate 
via nucleon-nucleon bremsstrahlung is calculated in a one-pion exchange model with
arbitrary nucleon degeneracy \cite{brinkmann,thompson_2000}. 
As a check to the calculation, the final distribution should have a characteristic 
temperature of the ambient matter with zero neutrino  chemical potential.

\begin{figure}[b]
\hbox to\hsize{\hfill\includegraphics{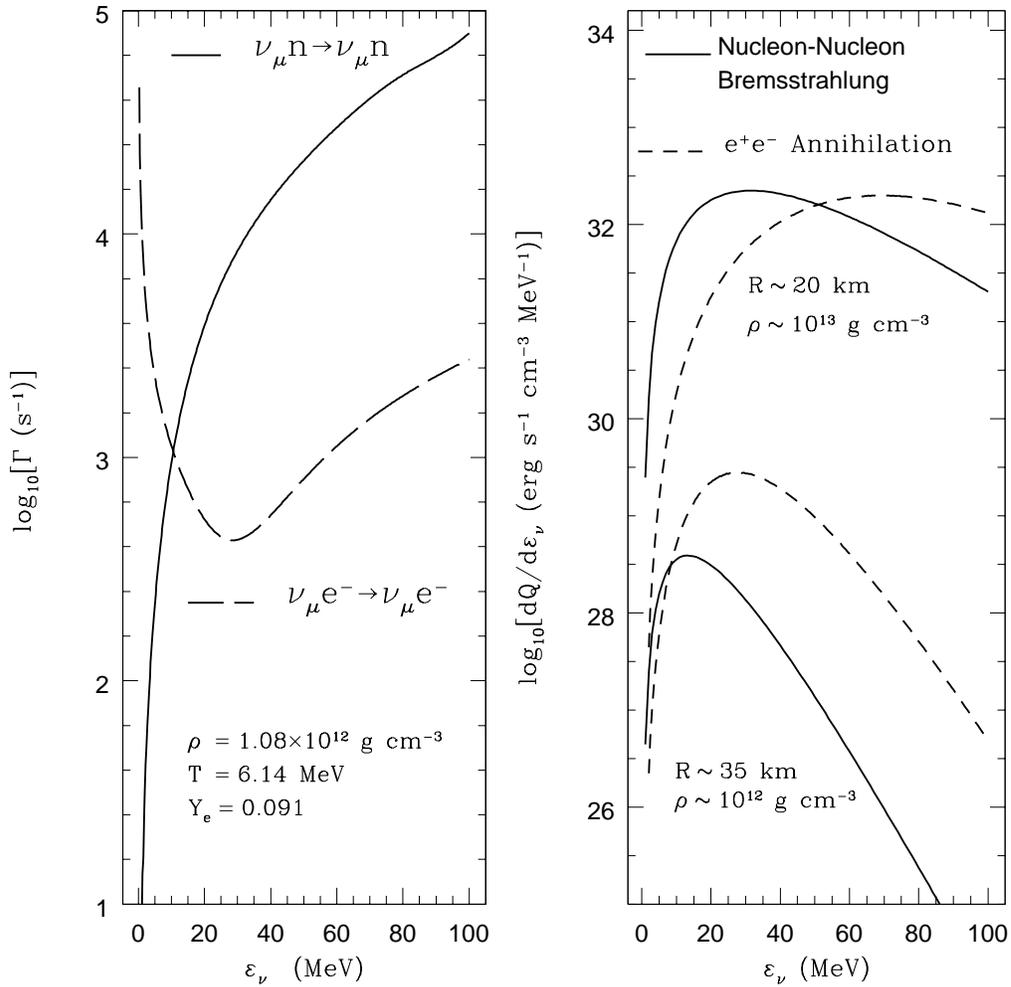}\kern+6in\hfill}
\caption{Left panel: the thermalization rate $\Gamma$ (s$^{-1}$) for both $\nu_\mu$-electron and $\nu_\mu$-neutron
scattering as a function of neutrino energy $\varepsilon_\nu$ (MeV) in equilibrium.  Right panel: the
differential emissivity versus $\varepsilon_\nu$ for both nucleon-nucleon bremsstrahlung
and $e^+e^-$ annihilation for two thermodynamic points taken from a representative
one-dimension core-collapse simulation \cite{bhf_1995}.}
\label{one}
\end{figure}

The left panel of Fig. \ref{one} shows the thermalization rates for $\nu_\mu n$ and $\nu_\mu e^-$
scattering, defined in terms of the average energy transfer ($\omega$) at that energy,
in equilibrium.  The calculation was performed with $T\simeq6.1$ MeV and 
$\rho\simeq1.1\times10^{12}$ g cm$^{-3}$.  These thermodynamic conditions are representative 
of the $\nu_\mu$ neutrinosphere, the semi-transparent regime where the neutrinos begin to
decouple from the matter and free-stream to infinity.  Most noticeable in this graph is the
fact that while $\nu_\mu e^-$ scattering dominates at low energies ($< 10$ MeV),
at modest and high energies $\nu_\mu n$ scattering competes with or dominates thermalization.
Throughout our calculations, at a variety of densities, temperatures, and compositions,
we find this behavior to be generic for the two scattering processes.  Indeed, the point
where $\nu_\mu n$ scattering begins to overwhelm $\nu_\mu e^-$ scattering seems always
to fall between approximately 10 and 20 MeV.

The right panel of Fig. \ref{one} reveals the same type of systematics for the production and absorption processes.
In this case, however, we plot the total differential emissivity for nucleon-nucleon bremsstrahlung
and $e^+e^-$ annihilation in equilibrium for two different thermodynamic points taken from a one-dimensional
core-collapse simulation \cite{bhf_1995,thompson_2000}.  Bremsstrahlung is clearly the dominant production mechanism
at low neutrino energies.  In the interior, at densities of order 10$^{13}$ g cm$^{-3}$, $e^+e^-$ annihilation 
begins to compete only at neutrino energies above 50 MeV.  In our time-dependent calculations, we find 
that bremsstrahlung always dominates production below 10-20 MeV at all points in a representative collapse
profile.

%-----------------------------------------------------------------------------------%

\section{Neutrino-Driven Protoneutron Star Winds}

A complete and self-consistent theory of the origin of all the elements has been the grand program of nuclear
astrophysics since the field's inception.  
The $r$-process, or rapid neutron capture process, 
is a mechanism for nucleosynthesis by which seed 
nuclei neutron capture on timescales shorter than those for $\beta^-$ decay.
With a sufficient neutron flux, capture continues to very neutron-rich isotopes and to the heaviest elements 
(e.g., Eu, Dy, Th, and U) producing unique abundance peaks at $A\sim80$, 130, and 195
\cite{burbidge,wallerstein,meyer_1994}.  
The $r$-process is only quelled when photodisintegration timescales approach 
those for neutron capture.  After the intense neutron flux lessens, $\beta^-$ decay populates the
primary stable isobar for a given atomic number.   
While the relevant nuclear physics is fairly well understood, the 
astrophysical site, which must exist in order to produce the elemental abundances we find in nature, 
is not known.
The viability of a site for $r$-process nucleosynthesis hinges on three characteristics: 
the asymptotic entropy per baryon ($s_f$), the electron fraction ($Y_e$), and the dynamical timescale 
($\tau_{\rm dyn}$).  
One proposed site is the protoneutron star wind that emerges after core collapse and shock reheating 
during a supernova \cite{bhf_1995,meyer_1992,woosley_hoffman}.  

Both numerical and analytic studies of the conditions in this neutrino-driven wind have been carried out previously
\cite{dsw,qian_woosley}.  Early calculations based on realistic supernova models produced interesting nucleosynthesis
and appreciable $r$-process yields, but overproduced nuclei near $N=50$ ($^{88}$Sr, $^{89}$Y, and $^{90}$Zr) 
\cite{woosley_94,takahashi_94}.  This problem was overcome by fine-tuning $Y_e$ in these 
simulations \cite{hoffman_1996a}, but no consensus on the other parameters (particularly, $s_f$) has yet been reached.
The pioneering analytic work of ref. \cite{qian_woosley} using simple wind models showed that entropies fell
short by a factor of $\sim2-3$ of those needed for the dynamic timescales and lepton fractions achieved.  
In other studies $s_f$ was artificially enhanced by a factor of $\sim5$ to achieve proper solar
$r$-process abundances \cite{takahashi_94}.  The fact that these calculations indicate an $s_f$ too low for 
$r$-process nucleosynthesis does not exclude the wind as a potential site.  A slight  increase in energy deposition
after the wind's acceleration phase \cite{qian_woosley} or the effects of general relativity have been shown 
to decrease $\tau_{\rm dyn}$ and increase $s_f$ \cite{cardall}, both
favorable to nucleosynthesis.  Recent general relativistic steady-state and hydrodynamical studies indicate 
that winds can generate all three $r$-process abundance peaks only when the protoneutron star is quite 
massive $\sim2\,M_\odot$ \cite{otsuki,sumiyoshi} and the total neutrino luminosity large ($\sim10^{52}$ erg s$^{-1}$).
These condition produce modest entropies ($\sim130$ baryon$^{-1}$ k$_{\rm B}^{-1}$), but short
$\tau_{\rm dyn}\sim6$ ms.  

The wind equations can be reduced to three ordinary coupled, critical 
differential equations for the evolution of temperature ($T$), mass density ($\rho$), and velocity
($v$) in radius.  $Y_e$ is held constant.  Solving these equations constitutes an eigenvalue problem.
The eigenvalue sought is the mass outflow rate $\dot{M}=4\pi r^2 \rho v$, or, alternatively,
the critical radius ($R_c$) where $v(R_c)=c_s$, the local speed of sound.  In practice, we impose two boundary
conditions at the protoneutron star surface ($R_o$), which we take to be the $\nu_e$ neutrinosphere:
(1) $T(R_o)=T_{\nu_e}$ and (2) $\tau_\nu(R_o)=-\int\kappa_\nu\rho\,dr=2/3$, where $\tau_\nu$ is the neutrino optical depth.
A third constraint must be imposed if we are to close the system of equations.  This we take to be the critical condition,
$v(R_c)=c_s$, which defines the outer radial boundary.  The system, now well-posed, can be solved using a relaxation
technique on an adaptive radial grid \cite{london}. The code then adjusts the
radial mesh in a Newton-Raphson sense in order to fulfill all boundary conditions simultaneously. Once the
critical point is determined, we use l'Hospital's rule to bridge it and then integrate to infinity using 
a simple Runga-Kutta scheme.  

\begin{figure}
\vspace{6.50in}
\hbox to\hsize{\hfill\includegraphics{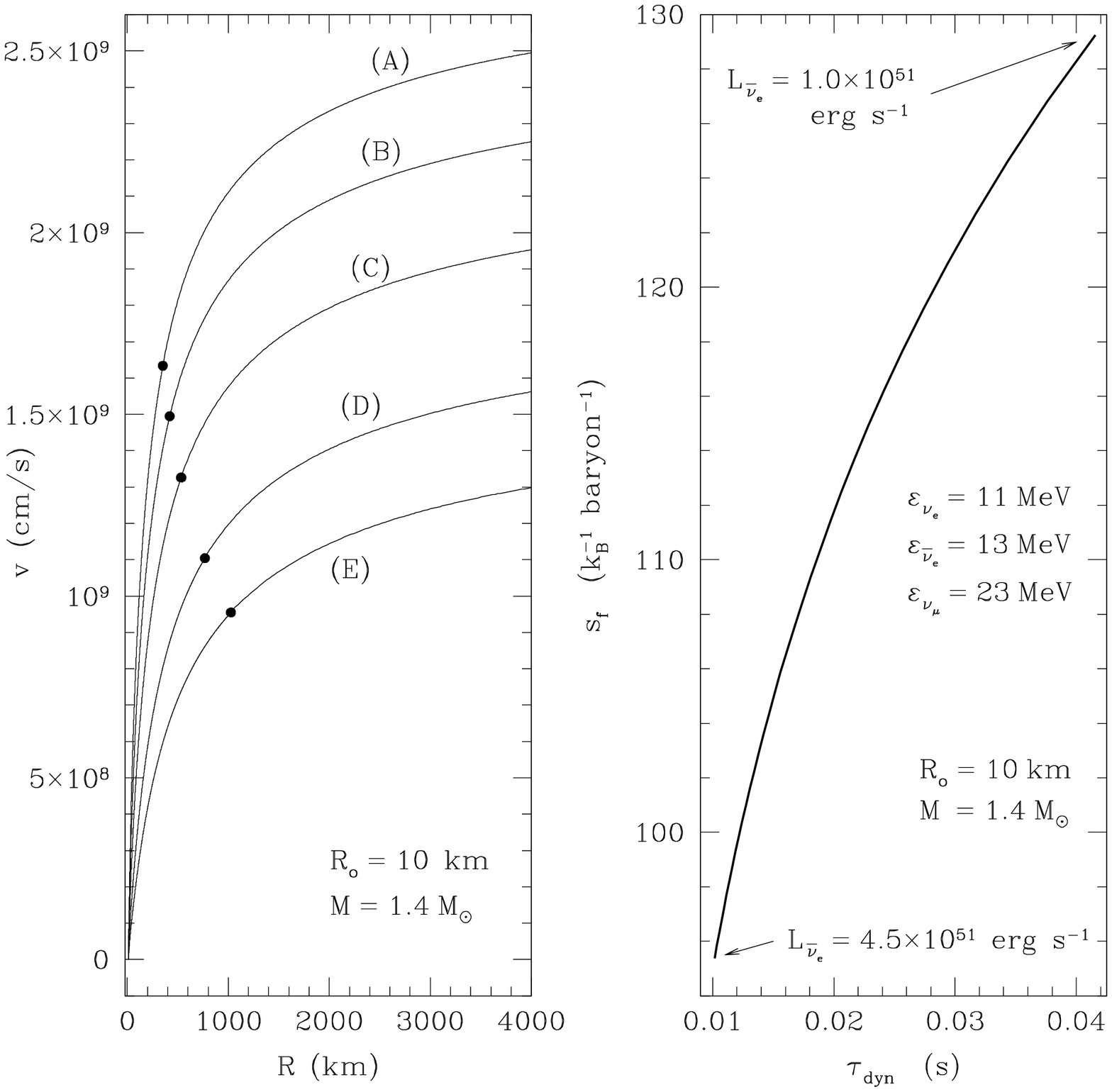}\kern+6in\hfill}
\caption{Left panel: general relativistic protoneutron star wind velocity $v$ (cm s$^{-1}$) profiles 
for six $\bar{\nu}_e$ luminosities: (A) $4.5$, (B) $3.5$, (C) $2.5$,
(D) $1.5$, and (E) $1\times10^{51}$ erg s$^{-1}$.
The mass and radius of the central neutron star are $M=1.4M_\odot$ 
and $R_o=10$ km, respectively.  The electron fraction ($Y_e$) is held constant 
at 0.302.  The dots denote the critical point where the 
velocity of the wind is equal to the local speed of sound. 
Right panel: asymptotic entropy ($s_f$) versus dynamical timescale ($\tau_{\rm dyn}$)
for constant protoneutron star mass and radius, 
but as a function of $\bar{\nu}_e$ luminosity.}
\label{two}
\end{figure}

The results of a preliminary and representative general relativistic calculation are shown in 
Fig. \ref{two}. The left panel shows velocity profiles for a variety of $\bar{\nu}_e$ 
luminosities, for a protoneutron star mass $M=1.4M_\odot$ and radius $R_o=10$ km.  This range
of neutrino luminosities is indicative of the first $\sim8-10$ seconds of the protoneutron star's life.
The right panel is a plot of the asymptotic entropy per baryon per Boltzmann constant versus 
$\tau_{\rm dyn}$ for the same protoneutron star.  Note that these calculations were carried out 
with constant $Y_e=0.302$.

In general, we find that higher entropies and shorter dynamical timescales result from the 
use of general relativity instead of Newtonian gravity.  For the 1.4 $M_\odot$ object we consider here, however,
we do not achieve entropies and timescales appropriate for $r$-process nucleosynthesis \cite{hoffman_1996a,otsuki,sumiyoshi}.

\section{Summary and Conclusions}

Our results for equilibration via $\nu_\mu$-electron scattering and
$\nu_\mu$-nucleon scattering demonstrate that the latter competes with 
or dominates the former as a thermalizer for neutrino energies 
$>10$ MeV for $\rho>1\times10^{11}$ g cm$^{-3}$ at all 
temperatures. At neutrino energies $>30$ MeV, the difference at 
all densities and temperatures is approximately an order
of magnitude. For the production and absorption processes, we find 
that nucleon-nucleon bremsstrahlung, at the average energy of an 
equilibrium Fermi-Dirac distribution at the local temperature, 
is 2 orders of magnitude faster than $e^+e^-$ annihilation at
$T\sim15$ MeV and  $\rho\sim10^{13}$ g cm$^{-3}$.  Only 
for $\rho\sim10^{12}$ g cm$^{-3}$ and $T\sim6$ MeV does 
$e^+e^-\leftrightarrow\nu_\mu\bar{\nu}_\mu$ begin to compete with 
bremsstrahlung at all energies.  We conclude from this study that 
the emergent $\nu_\mu$ and $\nu_\tau$ spectrum is (1) brighter and 
(2) softer than previously estimated \cite{burrows_1999}.  The former results from the 
inclusion of the new pair emission process, nucleon-nucleon bremsstrahlung.
The latter is a consequence of both the increased energy coupling 
between the nuclear and neutrino fluids through $\nu_\mu$-nucleon 
scattering and the fact that bremsstrahlung dominates $e^+e^-$ 
annihilation near the neutrinospheres at the lowest neutrino energies.

In addition, our first step in the calculation of realistic 
protoneutron star wind models has been successful; we have created
a robust technique for solving the steady-state eigenvalue problem
and have confirmed the results of other researchers qualitatively.
We plan to explore the parameter space of
protoneutron star winds exhaustively with an eye toward 
implementing full neutrino transport in a hydrodynamic simulation of 
wind emergence and evolution.  

\section{Acknowledgments}

The authors would like to thank the Nuclei in the Cosmos, 2000, organizing committee.
T.A.T. acknowledges the support of a NASA GSRP grant.


\begin{thebibliography}{9}
\bibitem{colgate} S. A. Colgate and R. H. White, Astrophys. J. {\bf 143}, 626 (1966). 
\bibitem{bethe} H. Bethe and J. R. Wilson, Astrophys. J. {\bf 295}, 14 (1985).
\bibitem{bhf_1995}A. Burrows, J. Hayes, and B. A. Fryxell, Astrophys. J., {\bf 450}, 830 (1995).
\bibitem{lamb_pethick} D. Lamb and C. Pethick, Astrophys. J., {\bf 209}, L77 (1976).
\bibitem{bruenn_1985}S. Bruenn, Astrophys. J. Supp., {bf 58}, 771 (1985).
\bibitem{fsb_1975}E. G. Flowers, P. G. Sutherland, and J. R. Bond, Phys. Rev. D, {\bf 12}, 2 (1975).
\bibitem{friman}B. L. Friman and O.V. Maxwell, Astrophys. J., {\bf 232}, 541 (1979).
\bibitem{reddy_1998}S. Reddy, M. Prakash, and J. M. Lattimer, Phys. Rev. D, {\bf 58}, 013009 (1998).
\bibitem{burrows_sawyer} A. Burrows and R. Sawyer, Phys. Rev. C, {\bf 58}, 554 (1998).
\bibitem{bs_1999}A. Burrows and R. Sawyer, Phys. Rev. C, {\bf 59}, 510 (1999).
\bibitem{reddy_1999} S. Reddy, M. Prakash, J. M. Lattimer, and J. A. Pons, Phys. Rev C {\bf 59}, 2888 (1999).
\bibitem{keil_1995}  W. Keil, H.-T. Janka, and G. G. Raffelt, Phys. Rev. D, {\bf 51}, 6635 (1995).
\bibitem{janka_1996} H.-T. Janka, W. Keil, G. Raffelt, and D. Seckel, Phys. Rev Lett. {\bf 76}, 2621 (1996).
\bibitem{raffelt_seckel} G. Raffelt and D. Seckel, Phys. Rev. Lett. {\bf 69}, 2605 (1998).
\bibitem{sigl_1997}G. Sigl, Phys. Rev. D {\bf 56}, 3179 (1997).
\bibitem{burrows_1999}A. Burrows, T. Young, P. Pinto, R. Eastman, and T. A. Thompson,  Astrophys. J., {\bf 539}, 865 (2000).
\bibitem{hannestad} S. Hannestad and G. Raffelt, Astrophys. J., {\bf 507}, 339  (1998).
\bibitem{brinkmann} R. P. Brinkmann and M. S. Turner, Phys. Rev. D, {\bf 38}, 8, 2338 (1988).
\bibitem{thompson_2000} T. A. Thompson, A. Burrows, and J. E. Horvath, Phys. Rev. C, {\bf 62}, 035802 (2000).
\bibitem{burbidge} E. M. Burbidge, G. R. Burbidge, W. A. Fowler, and F. Hoyle, Rev. Mod. Phys. {\bf 29}, 547 (1957).
\bibitem{wallerstein} G. Wallerstein, I. Iben, P. Parker, A. M. Boesgaard, G. M. Hale, A. E. Champagne, C. A. Barnes, 
	F. K\"{a}ppeler, V. V. Smith, R. D. Hoffman, F. X. Timmes, C. Sneden, R. N. Boyd, B. S. Meyer, and D. L. Lambert, 
	Rev. Mod. Phys. {\bf 69}, 995 (1997).
\bibitem{meyer_1994} B. S. Meyer, Annu. Rev. Astron. Astrophys. {\bf 32}, 153 (1994).
\bibitem{meyer_1992} B. S. Meyer, W. M. Howard, G. J. Mathews, S. E. Woosley, and R. D. Hoffman, Astrophys. J. {\bf 399}, 656 (1992).
\bibitem{woosley_hoffman} S. E. Woosley and R. D. Hoffman, Astrophys. J. {\bf 395}, 202 (1992).
\bibitem{dsw} R. C. Duncan, S. L. Shapiro, and I. Wasserman, Astrophys. J. {\bf 309}, 141 (1986).
\bibitem{qian_woosley} Y.-Z. Qian and S. E. Woosley, Astrophys. J. {\bf 471}, 331 (1996).
\bibitem{woosley_94} S. E. Woosley, J. R. Wilson, G. J. Matthews, R. D. Hoffman, and B. S. Meyer, Astrophys. J. {\bf 433}, 229 (1994).
\bibitem{takahashi_94} K. Takahashi, J. Witti, and H.-T. Janka, Astron. Astrophys. {\bf 286}, 857 (1994).
\bibitem{hoffman_1996a} R. D. Hoffman, S. E. Woosley, G. M. Fuller, and B. S. Meyer, Astrophys. J. {\bf 460}, 478 (1996).
\bibitem{cardall} C. Y. Cardall and G. M. Fuller, Astrophys. J. {\bf 486}, L111 (1997).
\bibitem{otsuki}K. Otsuki, H. Tagoshi, T. Kajino, and S. Wanajo, Astrophys. J., {\bf 533} (2000).
\bibitem{sumiyoshi} K. Sumiyoshi, H. Suzuki, K. Otsuki, M. Terasawa, and S. Yamada, astro-ph/9912156 (1999).
\bibitem{london}R. A. London and B. P. Flannery, Astrophys. J., {\bf 258} (1982).
\end{thebibliography}
\end{document}